\documentstyle[aps,prl,epsf,floats]{revtex}

\def\etal{{\it et al}}

\newcommand{\dst}{\displaystyle}
\newcommand{\be}{\begin{equation}}
\newcommand{\ee}{\end{equation}}

\newcommand{\txthalf}{{\textstyle{\frac{1}{2}}}}

\newcommand{\half}{{\frac{1}{2}}}
\newcommand{\csch}{{\mathrm{csch}}}

\begin{document}

\wideabs{ 
\title{Frenkel-Kontorova Model of Vacancy-Line Interactions on Ga/Si(112): Formalism}

\author{S. C. Erwin,$^{1,}$\cite{corresponding} A. A. Baski,$^2$ 
L. J. Whitman,$^1$ and R. E. Rudd$^{3}$}
\address{$^1$Naval Research Laboratory, Washington, D.C. 20375}
\address{$^2$Department of Physics, Virginia Commonwealth University,
Richmond, Virginia 23284}
\address{$^3$Department of Materials, University of Oxford, Oxford OX1
3PH, United Kingdom}

\date{\today}
\draft
\maketitle
\begin{abstract}
We describe in greater detail the exactly solvable microscopic model
we have developed for analyzing the strain-mediated interaction of
vacancy lines in a pseudomorphic adsorbate system (Phys. Rev. Lett.,
to appear).  The model is applied to Ga/Si(112) by extracting values
for the microscopic parameters from total-energy calculations. The
results, which are in good agreement with experimental observations,
reveal an unexpectedly complex interplay between compressive and
tensile strain within the mixed Ga-Si surface layer.
\end{abstract}

%
%
%
}

When a material is grown pseudomorphically on a lattice-mismatched
substrate, the resulting strain field can lead to self-organized
structures with a length scale many times the atomic spacing.  One
well known example is the Ge/Si(001) dimerized overlayer system. The
Ge film is compressively strained (by 4\% relative to the bulk), and
the system lowers its energy by creating dimer vacancies in the
surface layer;  at the vacancy sites, the exposed atoms in the second
layer rebond to eliminate their dangling bonds.  The missing-dimer
vacancies order into vacancy lines (VLs) with $2\times N$ periodicity,
where the optimal $N$ depends on the Ge coverage \cite{liu96}. Even
for coverages as low as a few monolayers, the concept of elastic
strain relaxation within a coherent pseudomorphic Ge film is
appropriate. For example, Tersoff showed theoretically for a 3-layer
Ge film that the equilibrium $N$ corresponds to the vacancy density at
which the compressive stress from the Ge overlayer cancels the tensile
stress from the rebonded missing dimers \cite{tersoff92}.

For monolayer and lower coverages, the concept of strain relaxation becomes
problematic, because the strain within a partial overlayer becomes
difficult to define. In this Letter we develop a model for
analyzing such situations and apply it to another VL system,
Ga on Si(112).  We show that despite striking similarities in the
phenomenology, the underlying energetics of Ga/Si(112) is rather
distinct from Ge/Si(001).  To make our treatment physically
transparent but also quantitatively accurate, we develop an exactly
solvable model of VL interactions in which the microscopic parameters
are extracted directly from first-principles total-energy
calculations.  This model contains only nearest-neighbor harmonic
interactions but reproduces the first-principles results quite
accurately, and thus allows for a particularly simple analysis of the
dominant interactions.

\begin{figure}[t]
\epsfxsize=6.5cm\centerline{\epsffile{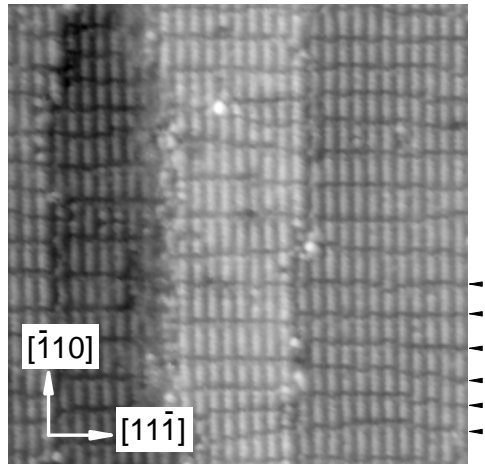}}
\caption{Empty-state STM image ($300\times 300$ \AA) of Ga/Si(112).
The wedges mark the vacancy lines, which are oriented along the
[11$\overline{1}$] direction (and are interrupted by a narrow
monolayer-high terrace in the
center of the figure).
\label{stm}}
\end{figure}

\begin{figure}[t]
\epsfxsize=7.0cm\centerline{\epsffile{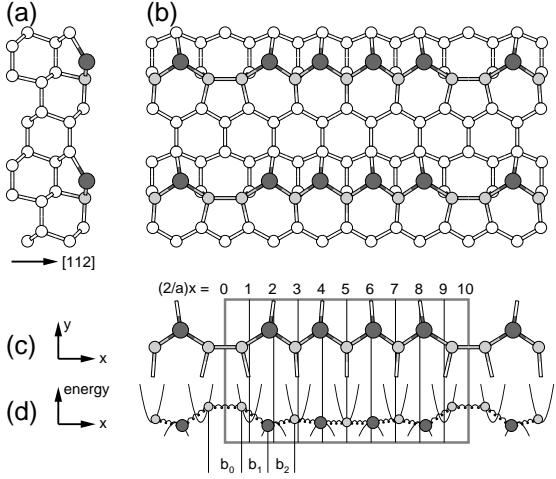}}
\caption{(a) Side and (b) top views of Ga/Si(112) with vacancy period
$N$=5. The fully relaxed coordinates are from first-principles
total-energy minimization. (c) Bonding chain of Ga (dark) and Si
(light) atoms. (d) One-dimensional Frenkel-Kontorova model
representing this Ga-Si chain. The (horizontal) atomic displacements,
$u_j$, are defined relative to the ideal positions (shown by thin grid
lines) of the substrate atoms; vertical displacements represent the
individual substrate-strain energies, $(1/2)k_j u_j^2$, appearing 
in Eq.~(\ref{FKenergy}).
\label{model}}
\end{figure}

When Ga is deposited on Si(112) and annealed, a well-ordered surface
is formed consisting of large (112)-oriented domains, as shown in
Fig.\ \ref{stm}. The VLs are oriented horizontally in Fig.\
\ref{stm}. Even at room temperature, the VLs show minimal thermal
meandering---only single kinks (up or down by one lattice spacing) are
observed in scanning tunneling microscopy (STM).  The mean VL spacing
in Fig.\ \ref{stm} is 5.2, with the distribution sharply peaked around
5 and 6.  The microscopic structure of VLs on Ga/Si(112) is shown in
Fig.\ \ref{model}.  Since Ga is trivalent it prefers to adsorb at
three-fold surface sites. The bulk-terminated Si(112) substrate, which
may be regarded as a sequence of double-width (111)-like terraces and
single (111)-like steps, offers just such three-fold sites at the step
edges, as shown in Fig.\ \ref{model}(b).  A single Ga vacancy leaves
two step-edge Si atoms exposed, which rebond to form a dimer. This
model was first proposed by Jung \etal. \cite{jung94} and subsequently
confirmed by Baski \etal. \cite{baski98} using STM and total-energy
calculations. It predicts that two adjacent Ga vacancies will be very
unlikely---because the three Si atoms exposed can form only one dimer,
resulting in an extra Si dangling bond; indeed, there are no adjacent
vacancies visible in Fig.\ \ref{stm}.  In this way, VLs on Ga/Si(112)
are essentially constrained to have a fixed width of one vacancy.

To construct a microscopic model for the interactions between VLs we
make three simplifying assumptions about their ground-state structure.
(1) We take the VLs to be perfectly straight; this turns a
two-dimensional surface problem into an effective one-dimensional
system. (2) We assume this one-dimensional system to be periodic with
a single vacancy per unit cell, that is, the vacancy separation is
taken to be $L=Na$, where $N$ is an integer and $a$ is the surface
lattice constant.  (3) Full structural relaxation within the
local-density approximation, described below, shows that the Si
substrate atoms are essentially unperturbed from their ideal locations
(Fig.\ \ref{model} shows the relaxed geometry for $N=5$; substrate
atoms are white).
Thus, we consider the vacancy-vacancy interaction to be mediated
entirely by the bonding chain of Si and Ga atoms shown in Fig.\
\ref{model}(c).

We now map this Ga-Si chain (with vacancy period
$N$) onto a one-dimensional chain of harmonic springs connecting $N$ Si atoms
and $N-1$ Ga atoms in a unit cell of length $Na$, as shown in Fig.\
\ref{model}(d).  The bonding of each atom, $j$, to the substrate is
also taken to be harmonic, with potential minima at the ideal
substrate positions.  This is a variant of the Frenkel-Kontorova (FK)
model, which has been widely used to study complex static and
dynamical phenomena arising from purely local interactions.  Within
the FK model, the total (potential) energy is
\begin{equation}
U = \sum_{i=0}^{2N-2} \frac{1}{2} K_i (b_i - b_i^{\rm eq})^2 +
\sum_{j=1}^{2N-1} \frac{1}{2} k_j u_j^2.
\label{FKenergy}
\end{equation}
Here the one-dimensional atomic displacements $u_j$ are defined with respect to the
ideal positions of the substrate atoms, $(a/2)j$, and the
corresponding spring lengths $b_i$ are defined in Fig.\
\ref{model}(d). 
This FK model has six parameters: within
each unit cell, one spring represents the Si-Si bond at the vacancy
(with spring constant $K_{\rm Si}$ and equilibrium length $b_{\rm
Si}^{\rm eq}$), while the remaining $2N-2$ springs represent the Ga-Si
bonds (with spring constant $K$ and equilibrium length $b^{\rm eq}$).
Two different substrate spring constants, $k_{\rm Si}$ and
$k_{\rm Ga}$, represent the bonds from Si and Ga atoms in the
chain to the rigid substrate.

To solve for the displacement field that minimizes the FK energy, we
first consider the force equations for atoms away from
the vacancies, and then apply the boundary conditions due to the
vacancies.  In equilibrium, the force on atom $j$ is given by
\be
F_j = 0 = K(2 u_j - u_{j-1} - u_{j+1} ) + k _j u_j ,
\label{basicForce}
\ee
which can be written for the Si or Ga sublattice as
\be
\left[ ( 2 K \! + \! k _{\rm Ga}) ( 2 K \! + \! k _{\rm Si}) - 2 k^2 \right] u_j =
K^2 \! \left( u_{j+2} \! + \! u_{j-2} \right) .
\label{subLattice}
\ee
In the continuum limit (large $N$) this becomes a simple differential equation,
\be
\left[\beta - 1\right] u(x) = (a/2)^2 \: u''(x),
\label{continuumLimit}
\ee
where we have defined
\be
\beta =  \left(1+\frac{k_{\rm Si}}{2K}\right) 
         \left(1+\frac{k_{\rm Ga}}{2K}\right).
\ee
Equation (\ref{continuumLimit}) shows that the continuum displacement field,
$u(x)$, has an exponential solution with the decay length
$(a/2)/\sqrt{\beta-1}$. Note that for the case of weak binding to the
substrate, $k_j \ll K$, this decay length reduces simply to
$\sqrt{K/k}\: (a/2)$, where $k$ is the average substrate potential.

For the discrete case (arbitrary $N$), Eq.~(\ref{subLattice}) provides
a recurrence relation whose solutions have the form
\be
u_j = c_1^{\rm Si,Ga} e^{2 \lambda j} + c_2^{\rm Si,Ga} e^{-2 \lambda j} ,
\label{discreteSolution}
\ee
where $e^{\pm 2 \lambda}$ are given by
\be
e^{\pm 2 \lambda }  =  \sqrt{\beta} \pm \sqrt{\beta - 1}.
\label{alpha}
\ee
The general form of the discrete solutions
is again seen to be exponential, although 
with a somewhat more complicated form for the (dimensionless) decay length,
$(2\lambda)^{-1}$. It is easy to verify by Taylor expansion that for 
weak substrate binding
we again recover the correct limiting behavior, 
$(2\lambda)^{-1} \rightarrow \sqrt{K/k}$. 

Eq.~(\ref{basicForce}) provides the following relationship between the
four coefficients $c^{\rm Si}_n$ and $c^{\rm Ga}_n$, which are used for 
the Si and Ga site respectively: 
\be
\frac{c_n^{\rm Ga}}{c_n^{\rm Si}} = 
\sqrt{\frac{1+k_{\rm Si}/2K}{1+k_{\rm Ga}/2K}}.
\label{Lambda}
\ee
It is convenient to introduce a dimensionless parameter, $\Lambda$,
defined by equating the rhs of Eq.~(\ref{Lambda}) with $\exp(2\Lambda)$.
The deviation of $\Lambda$ from zero measures the relative difference
in strength between Si- and Ga-substrate binding, and will give rise
(see below) to oscillations in the strain field around the simple
exponential behavior found in the continuum limit.

The remaining unknowns are determined by the boundary conditions at
the vacancies.  By applying these we arrive at the exact closed-form
solution for the displacement field and corresponding bond
lengths. For example the Si-Si vacancy-bond length, as a function of
vacancy period, can be written as
\be
b_0(N) = 
a - \frac{ 2 u_1^{\infty} }{  1 +\xi \left\{\coth [ 2(N-1)\lambda ]-1\right\}  },
\label{vacBond} \\
\ee
where $u_1^{\infty}$ and $\xi$ (both positive) are combinations of the various
FK parameters. The Ga-Si bond lengths have a more complicated form
due to the two different substrate potentials:
\begin{eqnarray}
b_i(N) & =  & {\dst
 \frac{a}{2} + 4 c \left\{ \cosh \Lambda \, \sinh \lambda
 \, \cosh [ (2i-2N+1) \lambda ] \right. }
\nonumber \\
 & &  {\dst \left.
 \pm \, \sinh \Lambda \, \cosh \lambda
 \, \sinh [ (2i-2N+1) \lambda ] \right\} },
\label{interiorBond}
\end{eqnarray}
where the last term (present only when $\Lambda\neq0$)
is positive and negative for
even- and odd-numbered
bonds, respectively.  The magnitude of the strain field is given by
the prefactor
$c =  {\dst (1/2) ( b_0 - a ) \, e^{-\Lambda } \, 
\csch \left[ 2 ( 1 - N ) \lambda \right] }$. It is evident from
Eqs.~(\ref{vacBond}) and (\ref{interiorBond}) that the strain field is
characterized by a single length scale, $(2\lambda)^{-1}$, which
describes both the relaxation of each bond with respect to the
distance, $i$, from a vacancy, and the relaxation of all strains with
respect to the vacancy period, $N$.

To apply this general solution to Ga/Si(112) we must determine
numerical values for the six FK parameters 
(two bond lengths, $b^{\rm eq}$ and $b_{\rm eq}^{\rm Si}$, and four spring
constants, $K$, $K_{\rm Si}$,  $k_{\rm Ga}$,  $k_{\rm Si}$) 
appearing in Eq.~(\ref{FKenergy}).  We do this either
analytically from the Stillinger-Weber potential (for Si-Si
parameters) or numerically from first-principles total-energy
calculations (for Ga-Si parameters).  These calculations were
performed in a double-sided slab geometry with six layers of Si
and a vacuum region equivalent to five layers of Si.  Total energies
and forces were calculated within the local-density approximation
(LDA) with gradient corrections
\cite{perdew92}, using Troullier-Martins pseudopotentials and a
plane-wave basis with a kinetic-energy cutoff of 8 Ry, as implemented
in the {\sc fhi96md} code \cite{bockstedte97}.  Total energies were
completely converged with respect to Brillouin-zone sampling. Full
structural relaxation was performed on all atoms except those in the
innermost double layer until the surface energies were converged to
0.1 meV/\AA$^2$.

Using the equilibrium structure of Ga/Si(112) with no vacancies we
find that, for small displacements, $k_{\rm Si}=2.0$ eV/\AA$^2$
and $k_{\rm Ga}=-0.7$ eV/\AA$^2$. The negative spring constant
here indicates that in the absence of defects the Ga sublattice is at
a point of unstable equilibrium---a finding confirmed by our LDA
results for finite $N$ (below). From calculations on isolated infinite
Ga-Si chains, we obtain $K=9.8$ eV/\AA$^2$ and $b^{\rm eq}=2.00$
\AA. This equilibrium bond length implies a 4\% compressive epitaxial
strain with respect to the Si substrate, and contributes to the driving
force for vacancy formation in Ga/Si(112). Finally, by expanding the
radial part of the Stillinger-Weber potential about the LDA vacancy
bond length for $N=2$, we obtain $K_{\rm Si}=5.5$ eV/\AA$^2$
and $b_{\rm Si}^{\rm eq}=2.25$ \AA.

\begin{figure}[t]
\epsfxsize=8.5cm\centerline{\epsffile{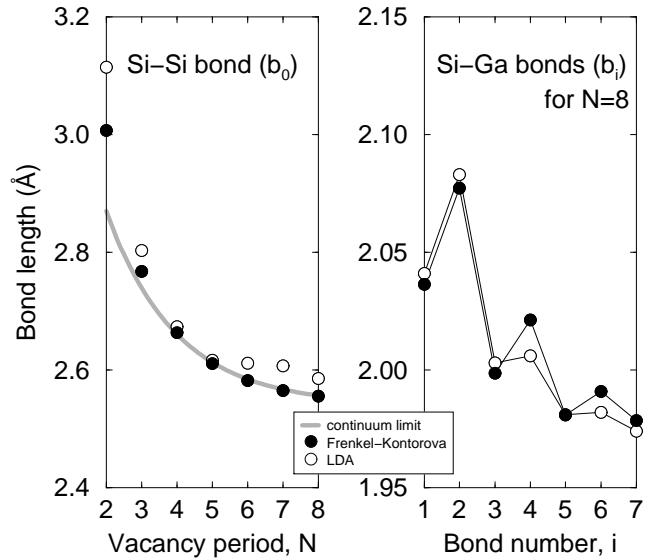}}
\caption{Equilibrium bond lengths from the FK model and LDA. 
Left: Si-Si vacancy bond length vs.\ vacancy period.  Right: Ga-Si
bond length vs.\ ``bond number'' [relative distance from the vacancy
bond, as defined in Fig.\
\protect\ref{model}(d)].
\label{b0bi}}
\end{figure}

We now compare the structural predictions of the FK model to those of
first-principles calculations. Using the LDA approach
described above, supercells with vacancy periodicities from $N=2$ to 8
were completely relaxed. Fig.\ \ref{b0bi} shows the FK and LDA results for the
vacancy bond length vs.\ $N$ and for the Ga-Si chain bond lengths for
$N=8$. The agreement is remarkably good, especially considering the
simplicity of the model. Note the oscillatory behavior of the
Ga-Si FK bond lengths---due to the unequal Ga- and Si-substrate
binding strengths---which is strikingly confirmed by the LDA results.

We turn next to the description of strain energetics and, thereby, the
effective VL interaction.  Within the FK model, we use Eqs.\
(\ref{vacBond}) and (\ref{interiorBond}) to evaluate the strain
energy, $U$, for arbitrary $N$. By exploiting the fact that the sums
over atomic sites are geometric series, we obtain a closed-form
solution for $U(N)$ (see Appendix). It is important to note that 
energies for different $N$ are not directly comparable, since the number
of Ga atoms per unit length varies with $N$. In general, this is resolved
by considering the surface free energy (per unit area),
\be
\gamma(N)=(NA)^{-1}\left[E_t(N)-(N-1)\mu_{\rm Ga}\right]/2, 
\label{surfaceEnergy}
\ee
where $E_t(N)$ is the (LDA) total energy for a Ga/Si(112) cell with vacancy
period $N$, and $\mu_{\rm Ga}$ is the Ga chemical potential
(the energy per atom of bulk Ga). The chemical potential is not
derivable within the FK model, but must be included ``by hand.'' We do
this by defining $\gamma^{\rm FK}(N)$ analogously to Eq.\
(\ref{surfaceEnergy}), with a fictitious chemical potential adjusted
to give identical VL formation energies, $\varepsilon_f$, within the
FK model and LDA (which gives $\varepsilon_f=-43$ meV/\AA).

\begin{figure}[t]
\epsfxsize=7.0cm\centerline{\epsffile{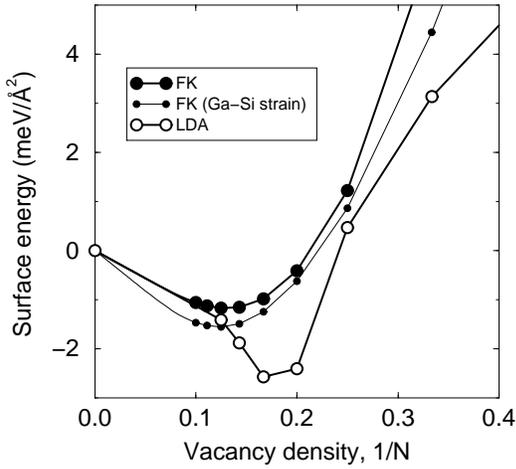}}
\caption{Relative surface energies, $\gamma(N)$, vs.\ vacancy density,
calculated within the FK model and LDA.
\label{esurf}}
\end{figure}

In Fig.\ \ref{esurf} we show the resulting FK and LDA surface
energies, relative to the limit of large $N$. The LDA energy minimum
occurs at $N=6$, in excellent agreement with the experimental results
for this system, while the FK minimum is only slightly higher, at $N=8$.  The
energy scale for the effective VL interaction is extremely small: relative to
infinite separation between vacancies, the LDA energy minimum is only of
order 2 meV/\AA$^2$, while the FK result is about half this value. In
general, the agreement between the two results is quite remarkable,
considering the length scale of the vacancy
spacing, the smallness of the energy scale, and the extreme simplicity
of the FK model.

By analyzing the individual contributions to the FK strain energy, it
is now simple to identify the dominant physical mechanism determining
the equilibrium VL spacing. In Fig.\ \ref{esurf} we plot (as small dots) only
those terms in Eq.~(\ref{FKenergy}) representing strains within the
Ga-Si bonding chain, and exclude the contribution from the Si-Si rebonded
dimer as well as all interactions between the chain and substrate
atoms. {\it It is clear that this Ga-Si chain strain completely
dominates the energetics for all physically relevant densities.}  This result
has two important implications. (1) The role of the rebonded Si
dimers, although obviously crucial for eliminating the dangling bonds
created by the Ga vacancies, plays no further significant energetic role in
determining the equilibrium VL spacing in Ga/Si(112). This is quite
different from the role of the rebonded Ge-dimer vacancies in
Ge/Si(001), for which the tensile stress contribution has been shown
to cancel the
compressive contribution from the overlayer only at the proper
density \cite{tersoff92}. (2) While a low density of Ga vacancies allows for relief of
compressive strain, when their density becomes too high, part of the
Si-Ga bonding chain experiences {\it tensile} strain---which acts as a
repulsive interaction between VLs. This effect is readily visible in the right
panel of Fig.\ \ref{b0bi}.  Relative to the equilibrium Si-Ga
bond length of 2.00 \AA, bonds far from the vacancy (numbered 5-7) are
compressively strained, bonds near the vacancy (3-4) are essentially
unstrained, but bonds very close to the vacancy (1-2) are tensilely
strained. This tensile contribution begins to dominate the surface 
energy at about $N<5$.

In summary, we have developed a Frenkel-Kontorova model to analyze the
microscopic origins of vacancy-line interactions on Ga/Si(112). The
model reveals that the mechanism of strain relaxation in this
submonolayer adsorbate system is quite complex. In particular, we have
identified the microscopic origins of attractive and repulsive
interactions between vacancy lines: both are mediated by a combination
of compressive and tensile bond strains within a single chain of Ga
and Si atoms. The sum of these strain energies is minimized at a
vacancy-line density very close to the experimentally observed
value. In general, we expect the future analysis of similar
strain-induced self-organized adsorbate systems to benefit from this
type of simple but accurate analytical model.

Computational work was supported by a grant of HPC time from the DoD
Major Shared Resource Center \mbox{ASCWP}\@.  This work was funded by
ONR\@. 


\appendix

\section*{Solution of the Frenkel-Kontorova Model}


\subsection{The Model}


Let the atoms be located at $x_j$ for $j=1,\ldots ,2N-1$.
Define the displacement by $u_j = x_j - (a/2)j$.
Then the potential energy is
\be
U = \sum _{i=0} ^{2N-2} \half K_i ( b_i - b_i^{eq} )^2 +
         \sum _{j=1} ^{2N-1} \half k_j \, u_j^2,
\label{energy}
\ee
\newpage
\onecolumn
\widetext
\noindent
where $K_i = K$ and $b_i^{\rm eq} = b^{\rm eq}$ for $i\ne 0$ and 
$K_0 = K_{\rm Si}$ and $b_0^{\rm eq} = b_{\rm Si}$. The bond lengths are
defined as
\be
b_i = \left\{ 
\begin{array}{cccc}
x_{i+1} - x_i & = & u_{i+1} - u_i + \frac{a}{2} 
& ~~~~~ {\mathrm for}~i=1,2,\ldots,2N-2\\
x_{1} - x_{-1} & = & u_{1} - u_{-1} + a 
& ~~~~~ {\mathrm for}~i=0.\\
\end{array}
\right.
\ee

\noindent
In the bulk (away from the vacancy) the force experienced by
an atom in equilibrium is

\begin{eqnarray}
F_j = 0 & = & K(2 u_j - u_{j-1} - u_{j+1} ) + k_j u_j \\
 & = & K \left[ \left( 2 + \frac{k_j}{K} \right) u_j - u_{j-1} 
- u_{j+1} \right].
\label{eom}
\end{eqnarray}

\noindent
At the vacancy, the force on the first Si atom is
\be
F_1 = 0 = K_{\rm Si} ( u_1 - u_{-1} + a - b_{\rm Si} ) - 
          K ( u_2 - u_1 + \frac{a}{2} - b^{\rm eq}) + k_{\rm Si} u_1.
\label{vacForce}
\ee

\noindent
Assuming periodicity, the chain is reflection symmetric about
$j=0$ and $j=N$, so that
\begin{eqnarray}
u_j & = & - u_{2N-j} 
\label{refSym} \\
u_1 & = & - u_{2N-1} = -u_{-1}.
\end{eqnarray}
Clearly we need only consider one end of the chain.
(\ref{vacForce}) then reduces to
\be
0 = (2 K_{\rm Si} + K + k_{\rm Si}) u_1 - K \, u_2 + (2 K_{\rm Si} - K) \frac{a}{2} 
      + ( K \, b^{\rm eq} - K_{\rm Si} \, b_{\rm Si} ).
\label{symmVacForce}
\ee


\subsection{The Displacements}


The bulk force equation (\ref{eom}) is linear, 
so the solution is a linear combination
of exponential terms such as

\be
u_j = c_j \, \alpha ^j.
\ee

\noindent
We plug this into (\ref{eom}) with the relation $c_j = c_{j-2}$,
which says that the Ga and Si displacements are scaled
differently, but that all of the displacements for either type of
atom are scaled by the same factor, that is,  
$c_j$ is $c_{\rm Si}$ for odd $j$ and $c_{\rm Ga}$ for even $j$.
The result is

\be
\left( 2 + \frac{k_j}{K} \right) c_j = 
\left( \alpha + \frac{1}{\alpha} \right) c_{j-1}.
\label{redSiteForce}
\ee

\noindent
First, we solve for $\alpha$.
Dividing the equation (\ref{redSiteForce}) on two adjacent sites and 
using $c_j = c_{j-2}$ to eliminate $\alpha$, we get

\be
\alpha + \frac{1}{\alpha} = 2 
\sqrt{\left(1+\frac{k_{\rm Si}}{2K}\right)
\left(1+\frac{k_{\rm Ga}}{2K}\right)  },
\label{coshlambda}
\ee
where $k_j$ is $k_{\rm Si}$ for odd $j$ and 
$k_{\rm Ga}$ for even $j$.
This may be solved for $\alpha$:
\begin{eqnarray}
\alpha _{\pm} & = & \sqrt{\beta} \pm \sqrt{\beta - 1} \\
\beta & = & \left(1+\frac{k_{\rm Si}}{2K}\right)
\left(1+\frac{k_{\rm Ga}}{2K}\right).
\label{alphaNotes}
\end{eqnarray}

\noindent
Note that $\alpha _+ = 1/\alpha _-$.  Also note that $\alpha$
could be complex for negative $k_j$.  Basically this says
that an inverted substrate potential can in principle cause the system to
become unstable.  If we assume $k_j \ll K$, which is true
for our system, then the system is stable as long as
$k_{\rm Si} + k_{\rm Ga}>0$.

Next, we solve for $c_j/c_{j-1}$. 
Multiplying the equation (\ref{redSiteForce}) on two adjacent sites and 
using $c_j = c_{j-2}$ to eliminate the $c_j$'s, we get

\be
\rho \equiv \frac{c_{\rm Ga}}{c_{\rm Si}} = 
\sqrt{\frac{1+k_{\rm Si}/2K}{1+k_{\rm Ga}/2K}}.
\label{rho}
\ee

\noindent
Thus, the displacements are given by
\be
u_j = \left\{ 
\begin{array}{cc}
\rho ^{-1/2} c_+ \, \alpha ^{j-N} + \rho ^{-1/2} c_- \, \alpha ^{N-j} 
& ~~~~~ {\mathrm for}~j~{\mathrm odd~({\rm Si}),}
\label{bulkDisp} \\
\rho ^{1/2} \, c_+ \, \alpha ^{j-N} + \rho ^{1/2} \, c_- \, \alpha ^{N-j} 
 & ~~~~~ {\mathrm for}~j~{\mathrm even~(Ga)}
\end{array}
\right .
\ee
with $\alpha = \alpha _+$ in (\ref{alphaNotes}) and $\rho$ given
by (\ref{rho}).


\subsection{The Vacancy Bond Length}


This leaves two undetermined coefficients, $c_+$ and $c_-$, which
are set by the boundary conditions at the vacancies.  As noted
above, the periodicity induces a reflection symmetry (\ref{refSym}),
so that
\be
c_+ = - c_- = c
\label{refCoeff}
\ee
and there is only one undetermined coefficient.
Substituting the bulk displacements (\ref{bulkDisp}) into
the vacancy force equation (\ref{symmVacForce}) we get
\be
c = \frac{ (K_{\rm Si} - \txthalf K) a + 
( K \, b^{\rm eq} - K_{\rm Si} \, b_{\rm Si})}{-\rho ^{-1/2} (2K_{\rm Si} + K + k_{\rm Si})
(\alpha ^{1-N} - \alpha ^{N-1}) + K \, \rho ^{1/2} (\alpha ^{2-N} - 
\alpha ^{N-2})} .
\label{coeff}
\ee

\noindent
Then the bond length at the vacancy is given by
\be
b_0 = 2 u_1 + a = 
 a +  \frac{ (2 K_{\rm Si} - K) a + 
2( K \, b^{\rm eq} - K_{\rm Si} \, b_{\rm Si})}{-(2K_{\rm Si} + K + k_{\rm Si})
+ K \, \rho  \left(
\frac{\alpha ^{2-N} - \alpha ^{N-2}}{\alpha ^{1-N} - \alpha ^{N-1}}
\right) }.
\label{vacBondLen}
\ee


\subsection{The Bond Lengths}


We can rewrite the displacements (\ref{bulkDisp}) in the form
\be
u_j = \txthalf c \left\{ \left[ (\rho ^{1/2} + \rho ^{-1/2}) +
 (\rho ^{1/2} - \rho ^{-1/2}) \cos (\pi j) \right] \,
(\alpha ^{j-N} - \alpha ^{N-j} )
\right\} .
\label{symDisp}
\ee

\noindent
The bulk bond lengths are computed from (\ref{bulkDisp}), 
(\ref{refCoeff}) and (\ref{coeff}):
\begin{eqnarray}
b_i & = & u_{i+1} - u_i + \txthalf a \\
& = & \frac{a}{2} + \frac{c}{2} \left[ 
\left( \rho ^{1/2} + \rho ^{-1/2} \right) \,
\left( \alpha ^{1/2} - \alpha ^{-1/2} \right) \,
\left( \alpha ^{i-N+1/2} + \alpha ^{N-i-1/2} \right) \right. \\
 & & \left. ~~~~- \cos (\pi i) \, 
\left( \rho ^{1/2} - \rho ^{-1/2} \right) \,
\left( \alpha ^{1/2} + \alpha ^{-1/2} \right) \,
\left( \alpha ^{i-N+1/2} - \alpha ^{N-i-1/2} \right) 
\right]
\nonumber \\
& = & \frac{a}{2} + 4 c \left[ \cosh \Lambda \, \sinh \lambda
 \, \cosh ( [2i-2N+1] \lambda ) \right. \label{bonds} \\
 & & ~~~~~~~~~ \left.
 - \cos (\pi i ) \, \sinh \Lambda \, \cosh \lambda
 \, \sinh ( [2i-2N+1] \lambda ) \right]
\nonumber 
\end{eqnarray}
where $\lambda = \half \log \alpha$ and $\Lambda = \half \log \rho$.


\subsection{The Energy}


Now that we have the expressions for the displacements (\ref{symDisp})
and the bond lengths (\ref{bonds}), they can be plugged into
(\ref{energy}) to compute the energy, $U$.
The following identity is useful to simplify the geometric
series that occur in the expression for the energy:
\be
\sum _{n=-N}^N x^n = \frac{x^{N+\half}-x^{-N-\half}}{x^{\half}-x^{-\half}}
\label{geomIdent}.
\ee

\noindent
Also it is useful to reexpress the displacements and couplings:
\begin{eqnarray}
u_j & = &  \txthalf c \left\{ \left[ (\rho ^{1/2} + \rho ^{-1/2}) +
 (\rho ^{1/2} - \rho ^{-1/2}) (-1)^j \right] \,
(\alpha ^{j-N} - \alpha ^{N-j} )
\right\}  
\label{symDisp2} \\
u_j^2  & = &
 \txthalf c^2 \left[
(\rho + \rho ^{-1} ) + (\rho - \rho ^{-1} ) (-1)^j 
\right] ( \alpha ^{2j-2N} + \alpha ^{2N-2j} - 2 ) \label{dispSq} \\
k_j & = & \txthalf \left[
( k_{\rm Ga} + k_{\rm Si} ) + ( k_{\rm Ga} - k_{\rm Si} ) (-1)^j
\right].  
\label{kappa2}
\end{eqnarray}

\noindent
Using (\ref{dispSq}) and (\ref{kappa2}), an expression for
the substrate bond energy may be derived by segregating the terms
into those that alternate in sign, and those that do not.  Then
the identity (\ref{geomIdent}) may be used to simplify the series:
\begin{eqnarray}
\sum _{j=1}^{2N-1} \txthalf k_j \, u_j^2 &=&
\frac{1}{2} c^2 \left( k_{\rm Ga} \rho + k_{\rm Si} \rho ^{-1} \right) 
\left[ -(2N-1) + \frac{\alpha ^{2N-1} - \alpha ^{1-2N}}{\alpha - \alpha ^{-1}}
\right]  \nonumber \\
& & ~~~~~ -
\frac{1}{2} c^2 \left( k_{\rm Ga} \rho - k_{\rm Si} \rho ^{-1} \right)
 \left[ -1 + \frac{\alpha ^{2N-1} + \alpha ^{1-2N}}{\alpha + \alpha ^{-1}}
\right] \\ [2mm]
& = &
\frac{1}{2} c^2 \left( k_{\rm Ga} \rho + k_{\rm Si} \rho ^{-1} \right) 
\left[ -(2N-1) + 
\frac{\sinh \left( 2 \lambda [2N-1] \right)}{\sinh \left( 2 \lambda \right)}
\right]  \nonumber \\
& & ~~~~~ -
\frac{1}{2} c^2 \left( k_{\rm Ga} \rho - k_{\rm Si} \rho ^{-1} \right)
 \left[ -1 + \frac{
\cosh \left( 2 \lambda [2N-1] \right)}{\cosh \left( 2 \lambda \right)}
\right] \\ [2mm]
& = &
2 c^2 K \left( \cosh 2 \lambda  - \cosh  2 \Lambda  \right) 
\left[ -(2N-1) + 
\frac{\sinh \left( 2 \lambda [2N-1] \right)}{\sinh \left( 2 \lambda \right)}
\right]  \nonumber \\
& & ~~~~~ +
2 c^2 K \, \sinh (2 \Lambda ) 
 \left[ -1 + \frac{
\cosh \left( 2 \lambda [2N-1] \right)}{\cosh \left( 2 \lambda \right)}
\right].
\label{substrateE}
\end{eqnarray}

\noindent
Now consider the bond energy.
Let
\begin{eqnarray}
a_0 & = & \txthalf a - b^{\rm eq} ,\\
c_1 & = & 4 c \, \cosh \Lambda \, \sinh \lambda ,\\
c_2 & = & 4 c \, \sinh \Lambda \, \cosh \lambda .
\end{eqnarray}

\noindent
Then the bond stretch (and its square) away from the vacancy are
\begin{eqnarray}
\delta b_i & = & b_i - b^{\rm eq}  = a_0 + 
c_1 \cosh \left( [ 2i - 2N + 1] \lambda \right) 
\nonumber \\
 & & ~~~~~~~
               - (-1)^{i} c_2 \sinh \left( [ 2i - 2N + 1] \lambda \right) 
\\[2mm]
\delta b_i^2 & = & a_0^2 + 
c_1^2 \left\{ \txthalf + 
\txthalf \cosh \left( 2 [ 2i - 2N + 1] \lambda \right) \right\} 
\nonumber \\
& & ~
+ c_2^2 \left\{ -\txthalf + 
\txthalf \cosh \left( 2 [ 2i - 2N + 1] \lambda \right) 
\right\} 
\nonumber \\
& & ~-  c_2 (-1)^i \left\{ c_1 \sinh \left( 2 [ 2i -2N+1] \lambda \right)
+ 2 a_0 \sinh \left( [ 2i -2N+1] \lambda \right) \right\}
\nonumber \\
& & ~
+ 2 a_0 c_1 \cosh \left( [ 2i -2N+1] \lambda \right). 
\label{stretchSquar}
\end{eqnarray}

\noindent
The following identities are useful:
\begin{eqnarray}
\sum _{n=1}^{2N} \cosh ( K \, [n - N - \txthalf ] ) & = &  
\frac{ \sinh ( N K )}{\sinh ( \half K )}\\
\sum _{n=1}^{2N} (-1)^n \sinh ( K \, [n - N - \txthalf ] ) & = &   
\frac{ \sinh ( N K )}{\cosh ( \half K )}.
\end{eqnarray}
They are essentially geometric series like (\ref{geomIdent}).
Also we use the following expressions:
\begin{eqnarray}
c_1^2 + c_2^2 & = & 8 c^2 \left[ \cosh ( 2 \Lambda ) \, \cosh ( 2 \lambda ) 
-1 \right] , \\
c_1^2 - c_2^2 & = & 8 c^2 \left[  \cosh ( 2 \lambda ) - \cosh ( 2 \Lambda ) 
\right] , \\
c_1 c_2 & = & 4 c^2 \sinh ( 2 \Lambda ) \, \sinh ( 2 \lambda ) , \\
\cosh ( \Lambda ) - \sinh ( \Lambda ) & = & e^{-\Lambda}.
\end{eqnarray}

\noindent
From (\ref{coshlambda}), we have
\be
\cosh ( 2 \lambda ) = \txthalf \left( \alpha + \frac{1}{\alpha} \right) = 
\sqrt{\left(1+\frac{k_{\rm Si}}{2K}\right)
\left(1+\frac{k_{\rm Ga}}{2K}\right)  }.
\ee

\noindent
The non-vacancy bond energy is then computed from (\ref{stretchSquar}):
\begin{eqnarray}
\sum _{i=1} ^{2N-2} \half K \, \delta b_i^2 & = & 
\txthalf K \left\{ \rule[-.9ex]{0mm}{4.2ex}
\txthalf (  2 a_0^2 + c_1^2 - c_2^2) (2N-2) + \right.
\nonumber \\
& & ~
\txthalf ( c_1^2 + c_2^2 )
\frac{\sinh \left( 4 \lambda [ N - 1] \right)}{\sinh \left( 2 \lambda \right)} 
+
\nonumber \\
& & ~  - c_1 c_2 
\frac{\sinh \left( 4 \lambda [ N - 1] \right)}{\cosh \left( 2 \lambda \right)} 
- 2 a_0 c_2 
\frac{\sinh \left( 2 \lambda [ N - 1] \right)}{
\cosh \left( \lambda \right)} +
\nonumber \\
& & ~ \left.
 2 a_0 c_1 
\frac{\sinh \left( 2 \lambda [ N - 1] \right)}{
\sinh \left( \lambda \right)} 
\right\} \\
& = & 
\txthalf K \left\{ \rule[-.9ex]{0mm}{4.2ex}
( [\txthalf a -b^{\rm eq}]^2 + 4c^2\, \left[ \cosh ( 2 \lambda ) - 
\cosh ( 2 \Lambda ) \right] ) 
(2N-2) + \right.
\nonumber \\
& & ~
4 c^2 \left[ \cosh ( 2 \Lambda ) \, \cosh ( 2 \lambda ) 
-1 \right] 
\frac{\sinh \left( 4 \lambda [ N - 1] \right)}{\sinh \left( 2 \lambda \right)} 
+
\nonumber \\
& & ~  -4 c^2 \, \tanh \left( 2 \lambda \right) \, 
\sinh \left( 2 \Lambda \right) 
\sinh \left( 4 \lambda [ N - 1] \right)
+ 
\nonumber \\ 
& & ~ \left. 8 (\txthalf a -b^{\rm eq}) c \, 
e^{-\Lambda} \, 
\sinh \left( 2 \lambda [ N - 1] \right) 
\rule[-.9ex]{0mm}{4.2ex}
\right\} .
\label{nonvac} 
\end{eqnarray}


\noindent
The vacancy bond energy is given by
\be
\txthalf K_{\rm Si} \delta b_0^2 =
\txthalf K_{\rm Si} \left\{
 (a - b_{\rm Si}) +  \frac{ (2 K_{\rm Si} - K) a + 
2( K \, b^{\rm eq} - K_{\rm Si} \, b_{\rm Si})}{-(2K_{\rm Si} + K + k_{\rm Si})
+ K \, \rho  \left(
\frac{\sinh (2\lambda [N-2])}{\sinh (2\lambda [N-1])}
\right) }
\right\} ^2.
\label{vacE}
\ee

\noindent
Finally,
\begin{eqnarray}
c & = &
 \frac{ (K_{\rm Si} - \txthalf K) a + 
( K \, b^{\rm eq} - K_{\rm Si} \, b_{\rm Si})}{2\rho ^{-1/2} (2K_{\rm Si} + K + k_{\rm Si})
\sinh (2 \lambda [N\! -\! 1]) - 2K \, \rho ^{1/2} 
\sinh ( 2 \lambda [N\! -\! 2])} \\
& = & 
 \frac{ \txthalf \, \rho ^{-1/2} \left\{ [(K_{\rm Si}/K) - \txthalf ] a + 
 [  b^{\rm eq} - (K_{\rm Si} /K)
\, b_{\rm Si}] \right\} }{
\rho ^{-1} (2(K_{\rm Si}/K) -1)\sinh (2 \lambda [N\! -\! 1]) 
 +   \sinh ( 2 N \lambda ) } ,
\label{finalc}
\end{eqnarray}
where we have used
\begin{eqnarray}
k_{\rm Ga} & = & 2K \left( \rho ^{-1} \cosh ( 2 \lambda ) - 1 \right) \\
k_{\rm Si} & = & 2K \left( \rho \,  \cosh ( 2 \lambda ) - 1 \right). 
\end{eqnarray}

\noindent
The energy, $U$, is the sum of (\ref{substrateE}), 
(\ref{nonvac}) and (\ref{vacE}),
using (\ref{finalc}) for $c$.


\subsection{Simplifications}


Below is a collection of simplified formulas.
We can rewrite (\ref{vacBondLen}) in the form of Eq.\ (\ref{vacBond}),
\be
b_0 = 
a - \frac{ 2 u_1^{\infty} }
{  1 +\xi \left\{\coth [ 2(N-1)\lambda ]-1\right\}  },
\ee
where 
\be
u_1^{\infty} = 
  \frac{ ( K_{\rm Si} - K/2) a + ( K \, b^{\rm eq} - K_{\rm Si} \, b_{\rm Si})}{
(2K_{\rm Si} + K + k_{\rm Si}) - K \, \rho  \, \alpha ^{-1} }
\ee

\noindent
and 

\be
\xi = \frac{K \, \rho \, \sinh ( 2 \lambda ) }{
-(2K_{\rm Si} + K + k_{\rm Si}) + K \, \rho  \, \alpha ^{-1} }.
\ee

\noindent
Further, we can express the coefficient $c$ in terms of $b_0$,
rather than vice-versa:

\be
c = \txthalf ( b_0 - a ) \, e^{-\Lambda } \, 
\csch \left( 2 [ 1 - N ] \lambda \right).
\ee


\begin{references}
\bibitem[*]{corresponding} On leave at:
Fritz-Haber-Institut, Faradayweg 4-6, D-14195 Berlin-Dahlem, Germany.
Electronic address: \mbox{erwin@dave.nrl.navy.mil}
\bibitem{liu96} F. Liu and M. Lagally, Phys. Rev. Lett. {\bf 76}, 3156 (1996).
\bibitem{tersoff92} J. Tersoff, Phys. Rev. B {\bf 45}, 8833 (1992).
\bibitem{jung94} T. M. Jung, S. M. Prokes, and R. Kaplan, J. Vac. Sci. Technol. A {\bf 12}, 1838 (1994).
\bibitem{baski98} A.A. Baski, S.C. Erwin, and L.J. Whitman, Surf. Sci. {\bf 423}, L265 (1999).
\bibitem{perdew92} J. P. Perdew {\it et al.}, Phys. Rev. B {\bf 46}, 6671 (1992).
\bibitem{bockstedte97} M. Bockstedte, A. Kley, and M. Scheffler, Computer Physics Commun. {\bf 107}, 187 (1997).
\end{references}
\end{document}